\documentstyle[12pt,aaspp4]{article}

\begin{document}

\title{The shape of the Red Giant Branch Bump as a diagnostic of 
partial mixing processes in low-mass stars}

\author{Santi Cassisi}
\affil{Osservatorio Astronomico di Collurania,
 via M. Maggini, 64100 Teramo, Italy; cassisi@te.astro.it.}
\author{Maurizio Salaris}
\affil{Astrophysics Research Institute, Liverpool John Moores 
       University, Twelve Quays House, Egerton Wharf, Birkenhead CH41 
       1LD, United Kingdom; ms@astro.livjm.ac.uk.}
\affil{Osservatorio Astronomico di Collurania,
 via M. Maggini, 64100 Teramo, Italy.}  
\affil{Max Planck Institut f\"ur Astrophysik,
Karl-Schwarzschild-Str. 1, 85748 Garching, Germany.}
\author{Giuseppe Bono}
\affil{Osservatorio Astronomico di Roma,
Via Frascati 33, 00040 Monte Porzio Catone, Italy; bono@coma.mporzio.astro.it.}

\received{}
\accepted{}

\pagebreak 
\begin{abstract}

We suggest to use the shape of the Red Giant Branch (RGB) Bump in metal-rich 
globular clusters as a diagnostic of partial mixing processes between the base 
of the convective envelope and the H-burning shell.  
The Bump located along the differential luminosity function of cluster RGB 
stars is a key observable to constrain the H-profile inside these structures. 
In fact, standard evolutionary models that account for complete mixing 
in the convective unstable layers and radiative equilibrium in the innermost 
regions do predict that the first dredge-up lefts over a very sharp 
H-discontinuity at the bottom of the convective region.  

Interestingly enough we found that both atomic diffusion and a moderate 
convective overshooting at the base of the convective region marginally affects 
the shape of the RGB Bump in the differential Luminosity Function (LF). 
As a consequence, we performed several 
numerical experiments to estimate whether plausible assumptions concerning 
the smoothing of the H-discontinuity, due to the possible occurrence of 
extra-mixing below the convective boundary, affects the shape of the RGB Bump.  
We found that the difference between the shape of RGB Bump predicted 
by standard and by smoothed models can be detected if the H-discontinuity 
is smoothed over 
an envelope region whose thickness is equal or larger than 0.5 pressure scale 
heights. Finally, we briefly discuss the comparison between theoretical 
predictions and empirical data in metal-rich, reddening free Galactic Globular 
Clusters (GGCs) to constrain the sharpness of the H-profile inside RGB stars.  
\end{abstract}


\pagebreak 
\section{Introduction}

One of the most intriguing features of GGCs is the occurrence of a local 
maximum in the distribution of RGB stars. It appears as a Bump in the 
differential LF, and as a change of slope in the cumulative LF. 
After the pioneering investigations by Thomas (1967) and Iben (1968), 
it is well known that this feature is due to the fact that during the 
RGB evolution the H-burning shell crosses the sharp chemical discontinuity 
left over by the convective envelope at the base of the RGB during 
its maximum sinking (first dredge-up). The abrupt change 
in the  hydrogen abundance affects the efficiency of the H-burning shell, 
since the stellar structure adjusts to the increase in the available fuel, 
and the stellar luminosity undergoes a temporary drop.
Dating back to the first detection in the GGC 47 Tucanae (King, Da Costa
\& Demarque 1985), the RGB Bump has been the crossroad of several
theoretical and observational investigations (Fusi-Pecci et al. 1990; 
Alongi et al. 1991; Bono \& Castellani 1992; Alves \& Sarajedini 1999; 
Ferraro et al. 1999; Zoccali et al. 1999; Bergbusch \& Vandenberg 2001, 
and reference therein).
The main reason for this effort is that the RGB Bump is a key observable 
to investigate the chemical profile inside RGB structures, and a robust 
diagnostic of the maximum depth reached by the outer convection during 
the first dredge-up.

Till a few years ago a quantitative comparison between theory and
observations was mainly hampered by the size of the available 
stellar samples along the RGB. This problem is particularly severe  
for the most metal-poor GGCs, since the RGB evolutionary timescales 
are shorter than in metal-rich ones. Moreover, the RGB Bump in 
metal-poor clusters is located at brighter magnitudes, and therefore 
in a RGB region poorly populated when compared with metal-rich 
clusters. However, the Hubble Space Telescope (HST) with its superior 
imaging quality and spatial resolution allowed us to firmly detect 
this feature in a sample of GGCs that cover a wide metallicity 
range (Zoccali et al. 1999).

This paper is the fifth in a series devoted to the RGB Bump in GGCs. 
Cassisi \& Salaris (1997) investigated the dependence of the Bump 
luminosity on the physical inputs adopted to construct  
stellar models. They showed that the difference in magnitude between 
the RGB Bump and the horizontal branch (HB) at the RR Lyrae instability 
strip ($\Delta V^{HB}_{Bump}$) predicted by their updated 
evolutionary models 
agrees quite well with empirical estimates of GGCs for which are available 
accurate spectroscopic measurements of cluster heavy element abundances.    
This evidence was further strengthened by the finding that 
$\Delta V^{HB}_{Bump}$ is only marginally affected by atomic diffusion 
(Cassisi, Degl'Innocenti \& Salaris 1997). A thorough comparison between 
predicted and observed $\Delta V^{HB}_{Bump}$ values  brought forth that 
theory and observations do agree at the level of $\approx0.1$ mag 
(Zoccali et al. 1999). To assess on a more quantitative basis the 
accuracy of current evolutionary models, Bono et al. (2001) compared 
the evolutionary lifetimes during the crossing of the H-discontinuity 
with the star counts across the RGB Bump. It turned out that theory 
is in very good agreement with observations over a wide metallicity 
range. Only a few clusters were at odds with theoretical predictions 
and 47 Tuc is one of them.

These investigations support the following evidence: 
i) the maximum depth reached by the convective envelope at the base 
of the RGB, whose diagnostic is the $\Delta V^{HB}_{Bump}$ parameter, 
is correctly predicted by stellar models. However, the parameter 
$\Delta V^{HB}_{Bump}$ relies on the distances provided by HB models. 
As a consequence, the agreement between theory and observations heavily 
relies on the adopted HB distance scale as well as on the GGC metallicity 
scales (Rutledge, Hesser \& Stetson 1997; Zoccali \& Piotto 2000; 
Bergbusch \& Vandenberg 2001);
ii) the extent of the H-discontinuity left over by the convective envelope, 
whose diagnostic is the star counts across the RGB Bump, does depend 
neither on the distance scale, nor on the metallicity scale and it is 
correctly predicted by theory (Bono et al. 2001).

The main aim of this investigation is to study in detail how the 
sharpness of the H-discontinuity affects the RGB Bump. 
In particular, we are interested in estimating how this jump in the 
chemical composition affects the shape of the Bump in the differential 
LF. The reason for such an investigation is twofold: 
i) current theoretical models do not firmly predict the efficiency 
and extent of all possible mixing processes occurring in stellar 
interiors as well as the occurrence of any partial mixing below the 
formal boundary of the convective region. It is worth mentioning that 
these physical mechanisms are no more pure speculative problems. 
In fact, Gratton et al. (2001) collected with VLT/UVES high dispersion 
spectra for a sizable sample of stars in two GGCs and found that atomic 
diffusion presents a very low efficiency when moving from MS to SGB stars
(see also Th\'evenin et al. 2001). This empirical evidence, once confirmed 
by new data, can supply useful hints on the occurrence of extra-mixing 
processes in the region located below the convective envelope. 
ii) The sharpness of the H-discontinuity does affect the evolutionary 
timescales (Bono \& Castellani 1992) of the H-burning shell during the 
crossing of the H-discontinuity, and in turn the shape of the Bump. 
Therefore a change in the extent of the smoothing region causes a change 
in the shape that can be constrained on the basis of the comparison with 
empirical data in metal-rich GGCs.

In \S 2 we briefly discuss the stellar models adopted in our
investigation, while in \S 3 we investigate how different 
assumptions on the smoothing lengths affect the shape of the Bump 
and the star counts across the Bump. This analysis is performed 
by using a Monte-Carlo technique. In this section we also discuss 
what are the observational requirements and the target that can 
allow us to detect a change in the shape of the Bump. 
Conclusions and future developments are outlined in \S 4.

\section{Stellar models}

The evolutionary code and input physics adopted in this investigation 
are the same as in Cassisi \& Salaris (1997). Theoretical models 
were transformed into the observational plane by adopting the bolometric 
corrections and color-temperature relations provided by Castelli, Gratton 
\& Kurucz (1997a,b).  We have focused our attention on a chemical composition 
typical of metal-rich GGCs, namely Y=0.23 and Z=0.006. Note that, 
if we account for an $\alpha$-enhancement of [$\alpha$/Fe]=0.4, this 
metallicity implies [Fe/H]=$-$0.8, i.e. the metallicity at which the number 
of RGB stars is expected to be the largest among the sample of GGCs.
The numerical experiments were also performed at fixed stellar mass 
$M/M_{\odot}=1$, since along the RGB the value of the evolving mass 
is almost constant. However, our conclusions do not depend on 
the selected mass value.  

We computed several series of evolutionary models, from the Zero Age
Main Sequence up to the RGB phases brighter than the Bump region. 
These models were constructed by neglecting or by accounting for atomic 
diffusion, and for various overshooting efficiencies namely $0.1-0.2 H_p$, 
where $H_p$ is the pressure scale height at the base of the convective 
envelope. According to Alongi et al. (1991) stellar models 
which include convective overshooting were constructed by assuming 
instantaneous mixing in the overshooting region. These numerical 
experiments were performed to estimate whether these two physical 
mechanisms affect the sharpness of the H-profile. 
Interestingly enough, we found that in all these models the H-discontinuity 
is sharp, and the shape of the Bump in the differential LF is the same as 
in standard models. This finding suggests that any detection of a peculiar 
shape in a cluster RGB Bump is almost certainly caused by a smoothing of 
the H-discontinuity.

As a consequence, we performed the experiments on our standard models, 
i.e. the evolutionary models that neglect both atomic diffusion and convective 
overshooting, and at the base of the RGB we artificially modified, according 
to Bono \& Castellani (1992), the abundance profiles below the point of 
maximum extent of the convective envelope.  
Then we evolved the new fictitious models well above the Bump region.
We have taken into account smoothing lengths of 0.1$H_{p}$,
0.2$H_{p}$, 0.5$H_{p}$, and 0.75$H_{p}$, where $H_{p}$ is the 
local pressure scale height at the H-discontinuity. For the sake of 
simplicity the chemical profiles in the smoothing region have been 
assumed to be linear, and the envelope abundances have been 
modified in such a way that the sum of element abundances by mass within 
the structure is perfectly conserved. Current smoothing lengths have been 
selected to fulfill two requirements: i) to avoid a large shift in 
the luminosity of the Bump along the LF. In fact, predicted and observed 
$\Delta V^{HB}_{Bump}$ are, within current observational and theoretical 
uncertainties, in satisfactory agreement. 
ii) To introduce a slight change in the sharpness of the H-discontinuity.
In fact, the hydrogen distributed in the smoothed region has to be 
taken from the chemical homogeneous region located above the H-discontinuity. 
However, substantial changes in the chemical composition 
(mainly the hydrogen abundance) of this region 
causes a sizable variation in the star counts across the Bump. Once again 
current theoretical predictions seem to agree quite well with empirical 
estimates (Bono et al. 2001).

\placefigure{fig1}

Figure~1 shows the hydrogen profile (H abundance per unit mass) around the
lower edge of the convective region at its maximum extent. The
dashed line shows the canonical profile and presents a sharp H-discontinuity,
while the solid line shows the H-profile after a linear smoothing of 
0.5$H_p$ has been applied.
The H abundance in the envelope attains quite similar values in the two 
cases, and indeed the difference is smaller than 0.001. This means that 
the jump in the H abundance between the envelope and the interior is 
very similar, but in the nonstandard models the H-discontinuity has been 
smoothed over a thicker region when compared with the standard one. 

\section{The effect of the H-discontinuity on the shape of the Bump}

To study in more detail the effect of the sharpness in the H-discontinuity 
on the shape of the RGB Bump, we computed the RGB luminosity functions of 
current stellar models. Figure 2 shows the LF of the RGB region located 
across the Bump for our standard models i.e. $M/M_{\odot}=1$, Z=0.006, 
Y=0.23 constructed by neglecting both atomic diffusion and convective 
overshooting. 

\placefigure{fig2}

The LF has been computed using a Monte Carlo technique, the number of
objects at a given luminosity being proportional to the local
evolutionary timescale. We have used an extremely large number of
objects to avoid spurious statistical fluctuations in a given 
luminosity bin. The bin size of the LF plotted in Fig.~2 is 0.02 mag.
It is worth mentioning that the intrinsic shape of the Bump
is asymmetric and presents a well-defined peak at its brightest end. 
Note that both the atomic diffusion and the convective 
overshooting below the formal boundary of the convective envelope 
do not change the shape of the Bump, since in these models the 
H-discontinuity is as sharp as in standard models.  

According to Bono et al. (2001) we define the parameter $\rm R_{Bump}$, 
as the number of stars located within $\pm$0.4 mag above and below the 
luminosity peak 
of the Bump, normalized to the number of RGB stars in the region between 
+0.5 and +1.5 mag below the peak of the Bump. The standard models supply  
$\rm R_{Bump}=$0.527, while the models constructed by accounting for 
atomic diffusion, convective overshooting, or the smoothed hydrogen 
profile have values within 0.05 of the standard one. This difference is 
quite negligible, since it is smaller than the typical observational 
error bar\footnote{The $R_{Bump}$ value of the smoothed models is similar 
to the standard one, since the current smoothing length was chosen to 
avoid a substantial change in the jump of the H-profile.} 
(see Bono et al. 2001).
On the other hand, the luminosity peak of the Bump shifts by 
$\sim$0.07 mag in the models that account for atomic diffusion,  
(see also Cassisi, Degl'Innocenti \& Salaris 1997), while for the 
models that include convective overshooting the brightness changes 
according to the following derivative $\sim$0.8 mag/$H_{p}$. 
At the same time, we found that an increase of 0.1$H_{p}$ in the 
smoothing length causes in the nonstandard models a shift of $\sim$0.025 
magnitudes in the position of the Bump. 
Figure~3 shows the effect of the smoothing on the shape of the Bump. 
The dashed line refers to the standard LF, while the solid lines display 
the LFs of smoothed models. They have been plotted by shifting the brightest 
boundary of the nonstandard Bumps over the standard one. The different 
LFs are normalized to the same number of stars above the Bump (where 
the hydrogen profile is the same in all models).

\placefigure{fig3}

The change in the shape of the Bump as a function of the smoothing 
length is quite evident. In particular, the Bump becomes more centrally 
peaked and more symmetric for an increase in the thickness of the smoothing 
region. For smoothing lengths equal or larger than 0.5$H_{p}$ the 
shape of the Bump is substantially different than for the standard 
models. According to the numerical experiments we already performed, 
such an effect cannot be produced by atomic diffusion nor by convective
overshooting.  
The reason for the difference in the shape of the Bump is outlined 
in Fig.~4. This figure shows the evolution in the 
Color-Magnitude-Diagram of our standard models and of the 
models with the H-discontinuity smoothed over a region of 0.5$H_{p}$. 
A glance at the data plotted in this figure clearly shows that the 
Bump region narrows in the latter case, and this change causes a 
narrowing of the Bump in the differential LF as well.

The physical explanation of this behavior has to be related to
the abrupt change of the mean molecular weight $\mu$ in the region across the 
chemical discontinuity left over by the outer convection during the first 
dredge up, and to the strong dependence of the H-burning efficiency 
on $\mu$. Standard models show a sharp change in the mean 
molecular weight at the chemical discontinuity; this feature strongly 
affects the H-burning efficiency, and in turn the stellar surface 
luminosity. In stellar models whose H-discontinuity has been smoothed, 
the sharpness of the $\mu$ variation is anticorrelated with the thickness 
of the smoothing region; as a consequence, the change of the H-burning 
efficiency is significantly lower in smoothed models than in canonical 
ones. The top panel of Fig.~5 shows the H-burning luminosity 
produced via CNO cycle as a function of time. 
Data plotted in this figure disclose that the luminosity drop 
taking place during the crossing of the H-discontinuity is clearly 
correlated with the smoothing length.   

The smoothing of the H-discontinuity affects not only $\mu$, but also 
the opacity profile within the structure. To test the hypothesis that the change of
opacity may play a role in causing the luminosity drop at the bump
location, we performed the following test. We constructed a fictitious
model in which the opacity in the region just below the H-discontinuity has 
been computed by adopting the same chemical composition of the 
region located above the H-discontinuity. Data plotted in the bottom 
panel of Fig.~5 clearly show that the change in the opacity profile 
caused by the variation in the chemical stratification has a negligible 
effect on the RGB Bump. Therefore, it is the change of $\mu$ which
determines the shape of bump region.

\placefigure{fig4}

At the same time, we have to account for the difference in the evolutionary 
timescales, since the number of stars in a given magnitude bin is proportional 
to this quantity. From top to bottom the panels in Fig.~6 show, for the two 
models plotted in Fig.4, the logarithm of the stellar counts along the three 
branches of the Bump region as a function of the V magnitude. The data for 
the nonstandard models were shifted in brightness as in Fig.~3. 
The arrows mark the direction during these evolutionary phases, i.e. the 
change in the surface luminosity. Panel a) shows the star counts before the 
temporary decrease in luminosity, while panel b) during the subsequent 
decrease in luminosity. The evolutionary timescale of the standard models  
at the bin where the luminosity does not increase is roughly a factor of 
2 longer than in the smoothed models, while it is systematically shorter  
during the phases approaching this point and the phases in which the 
luminosity decreases. During this latter phase the standard models reach  
the secondary minimum in luminosity quite rapidly when 
compared with the smoothed one. In fact, the evolutionary timescale of the 
smoothed models in the fainter luminosity bin is a factor of 3 longer 
than in the standard models. 
Panel c) shows the predicted stellar counts during the 
phases in which the surface luminosity increases once again. The evolutionary 
timescale of the standard models is essentially constant, while the smoothed 
models evolve more rapidly until they approach the bright edge of the 
Bump region.

\placefigure{fig5}

In this section we have discussed the difference in the LF between standard 
and smoothed models, now we wish to address the following question:
can this difference be detected in actual RGB Bumps?
To answer this question we performed new Monte-Carlo simulations,
by accounting for different sample sizes and observational errors.
Fig. 7 shows the same LFs plotted in Fig. 2, but they were computed 
by adopting a bin size of 0.05 mag and a 1$\sigma$ random photometric 
error (the accuracy of the photometric zero point is not relevant in 
this discussion) by 0.025 mag.
Data plotted in this figure suggest that smoothing lengths equal or 
larger than 0.5$H_{p}$ still affect the shape of the Bump in such a way 
that the difference between standard and smoothed models can be detected. 
According to our simulations, to detect smoothing lengths equal or 
larger than 0.5$H_{p}$ in metal-rich clusters with metal abundances larger  
than [Fe/H]=$-$1 and not affected by differential reddening, are necessary 
stellar samples larger than $\sim$120 stars within $\pm$0.2 mag the peak of 
the Bump, a bin size not larger than 0.06 mag and a 1$\sigma$ random 
photometric errors not larger than 0.03 mag.

\placefigure{fig6}

For metallicities equal or lower than [Fe/H]=$-$1 the change in the shape 
of the Bump can be barely detected. Fig. 8 shows the standard RGB LF for 
Z=0.006 (dotted line) together with the RGB LF obtained from a 1$M_{\odot}$ 
standard models (no diffusion, no overshooting) with Z=0.002 (solid line). 
By assuming an $\alpha$-enhancement [$\alpha$/Fe]=0.4 the latter metallicity 
becomes [Fe/H]=$-$1. The dashed line refers to the RGB LF of evolutionary 
models constructed by adopting the same mass value and metallicity 
(1$M_{\odot}$, Z=0.002) but with the H-discontinuity smoothed over 
0.5$H_{p}$. The LFs of the less metal-rich models have been shifted 
by 0.5 mag to match the bright edge of the Bump region of the LF 
for Z=0.006.

\placefigure{fig7}

Even though the main effect of the smoothing is qualitatively similar 
to the case of more metal-rich models, the extent of the Bump for Z=0.002 
is smaller both in the standard and in smoothed models. This means that 
smaller photometric errors, smaller bin sizes, and larger samples of 
RGB stars are mandatory to detect a change in the shape of the Bump.   
This result seems in contrast with the finding brought out by 
Bono et al. (2001) that the value of the $\rm R_{Bump}$ parameter 
is almost constant with metallicity. However, the contradiction is only 
apparent, and indeed toward lower metallicities the Bump becomes brighter, 
and therefore the number of stars in the normalization region decreases  
as well. This effect makes the $\rm R_{Bump}$ value only marginally 
dependent on the cluster metallicity.

\placefigure{fig8}

\section{Conclusions and final remarks}

We investigated in detail the dependence of both shape and location 
of the Bump along the RGB LF on the smoothing of the H-discontinuity
at the base of the convective envelope in Population II structures. 
The reason to account for this smoothing is related to the possible
occurrence of partial mixing processes below the lower boundary 
of the convective region. Current spectroscopic measurements of RGB 
stars are somehow at odds with theoretical predictions. To account for 
the low carbon isotopic ratios and for the Li abundance in population II
stars it has been suggested that rotation-induced mixing occurs above  
the RGB Bump (Sweigart \& Mengel 1979; Charbonnel 1995). However, even 
by accounting for this extra-mixing process the agreement between empirical 
data and theory is far from being satisfactory (Th\'evenin et al. 2001).  
The occurrence of mixing below the bottom of the convective region has 
been generally neglected, since theoretical predictions (Mestel 1957) 
and spectroscopic data  (Charbonnel, Brown, \& Wallerstein 1998, 
and references therein)
suggest that the occurrence of gradients in the molecular weight, such as 
the H-discontinuity in the envelope of RGB stars, stabilizes the mixing 
processes induced by rotation. However, current spectroscopic measurements  
support the evidence that RGB stars located across the Bump present large 
Li abundances (Charbonnel \& Balachandran 2000). This suggests that these
stars underwent Li production soon before or at the Bump phase.
The comparison between predicted and observed shape of RGB Bumps can 
supply tight constraints on the sharpness of the H-discontinuity, and 
in turn on the inhibiting effect of molecular weight barriers on 
extra-mixing related either to rotation or to exotic phenomena.
   
We constructed several stellar models by assuming different smoothing 
lengths, and we found that for smoothings up to 0.75$H_{p}$ current 
theoretical predictions concerning the difference in luminosity 
between the peak of RGB Bump and the HB, $\Delta V^{HB}_{Bump}$, and 
the star counts across the Bump region, $R_{Bump}$, are still in 
satisfactory agreement with observations. However, as expected, a change 
in the H-abundance profile significantly affects the evolutionary 
timescales of the H-burning shell, and in turn the shape of 
the RGB Bump along the LF. 
Interestingly enough, we found that it is possible to constrain 
the sharpness of the H-discontinuity at the base of the envelope 
of RGB stars by comparing predicted and empirical Bump shapes of  
differential LFs in metal-rich GGCs. 

In this context it is worth mentioning that this finding is far 
from being a speculative issue. In fact, the new Advanced Camera 
for Surveys (ACS, Clampin et al. 2000) on board of HST will supply, 
thanks to its wide field of view ($202\times202$ arcsec$^2$) and homogeneous 
spatial resolution (0.05 arcsec/pixel), a complete census of RGB stars over a 
substantial fraction of GGCs innermost regions. In particular, for clusters 
characterized by relatively high central densities, the ACS can measure at the 
star-by-star level the bulk of cluster RGB stars. 
As a consequence, the detection of the shape of RGB bump among 
massive metal-rich clusters will become feasible in the near 
future.  To constrain on a quantitative basis the empirical requirements 
necessary to supply an accurate detection of the shape of the Bump.  
we performed several Monte-Carlo experiments according to current 
theoretical predictions. We found that a robust detection requires 
a minimum sample of more than $\sim$120 RGB stars within $\pm0.2$ mag the 
peak of the Bump, the use of a bin size up to 0.06 mag in the 
LF, as well as random photometric errors smaller than 0.03 mag.  
Fortunately enough, several metal-rich GGCs and the new scientific 
capabilities of HST can allow us to cope with this challenging and 
intriguing project.

We are grateful to an anonymous referee for useful suggestions that 
improved the content of an early draft of the paper.
M.S. wishes to acknowledge the Osservatorio Astronomico di Collurania and 
the Max Planck Institut f\"ur Astrophysik for their kind hospitality during 
the completion of this work.
We warmly thank M. Zoccali for several interesting discussions on 
HST photometry and for useful comments on early draft of this paper.
This work was supported by MURST/Cofin2000 under the project: "Stellar 
Observables of cosmological relevance" (S.\ C. \& G. \  B.).

\pagebreak 

\pagebreak 
\figcaption{
Hydrogen profile (abundance by mass fraction) as a function of the
stellar mass fraction. The dashed line shows the H-profile of our standard 
models (no atomic diffusion, no convective overshooting) constructed by 
adopting $M/M_\odot=1.0$, Y=0.23, and Z=0.006, while the solid line 
refers to models constructed  with same input parameters but the 
H-discontinuity was smoothed out over a region whose thickness is 0.5$H_{p}$. 
See text for more details. 
\label{fig1}}

\figcaption{
Red giant branch differential LF, i.e. logarithm of star counts, as a
function of $M_{V}$ across the Bump region of the standard models. The 
bin size is of 0.02 mag.
\label{fig2}}

\figcaption{
Comparison between the differential LF of our standard models (dotted 
line) and the LFs of the models constructed by artificially smoothing 
the H-discontinuity over a region of 0.1$H_{p}$, 0.2$H_{p}$, 0.5$H_{p}$,
0.75$H_{p}$ (solid lines). The LFs of the smoothed models have been shifted 
in luminosity to match the bright edge of the Bump of the standard models.
\label{fig3}}

\figcaption{
Color-Magnitude-Diagram of the RGB Bump phase for the standard models  
(dotted line) and for the models with the H-discontinuity smoothed out 
over a region of 0.5$H_{p}$ (solid line).
\label{fig4}}

\figcaption{
{\sl Top}: Time behavior of the H-burning luminosity
produced via CNO cycle during the RGB Bump phase for our standard
model and models with artificially smoothed H-discontinuity. 
{\sl Bottom}: Time behavior of the stellar surface luminosity 
during the RGB Bump phase for a standard model (solid line), 
a model constructed by smoothing the H-discontinuity over a 
region of 0.5$H_{p}$ (long dashed line), and a standard model 
computed by artificially changing the opacity in the region 
across the H-discontinuity (dotted line, see text for more 
details).
\label{fig5}}

\figcaption{
Comparison between the differential LFs of the standard models (dotted line) 
and of the models whose H-discontinuity whose smoothed out over a region of 
0.5$H_{p}$ (solid line). From top to bottom the three panels refer to 
different evolutionary phases across the bump, namely  {\em a)} from the 
faint edge of the Bump to the secondary maximum,  {\em b)} from the secondary 
maximum to the secondary minimum, and {\em c)} from the secondary minimum to 
the bright edge of the Bump. The arrows mark the evolutionary direction, 
i.e. the change in the surface luminosity along the three branches.
\label{fig6}}

\figcaption{
Same as in Fig.~3, but the differential LF was constructed by adopting a 
bin size of 0.05 mag and 1$\sigma$ random photometric errors of 0.025 mag.
\label{fig7}}

\figcaption{
Comparison between different differential LFs constructed by adopting the same 
stellar mass -$M/M_\odot=1.0$- but different chemical compositions. The dotted 
and the solid line show the LFs of the standard models constructed by adopting 
Z=0.006 and Z=0.002 respectively.  The dashed line displays the LF of the model 
with Z=0.002 and the H-discontinuity smoothed out over a region of 0.5$H_{p}$. 
The bin size is of 0.02 mag. The LFs for Z=0.002 have been shifted in 
luminosity to match the bright edge of the Bump. 
\label{fig8}}

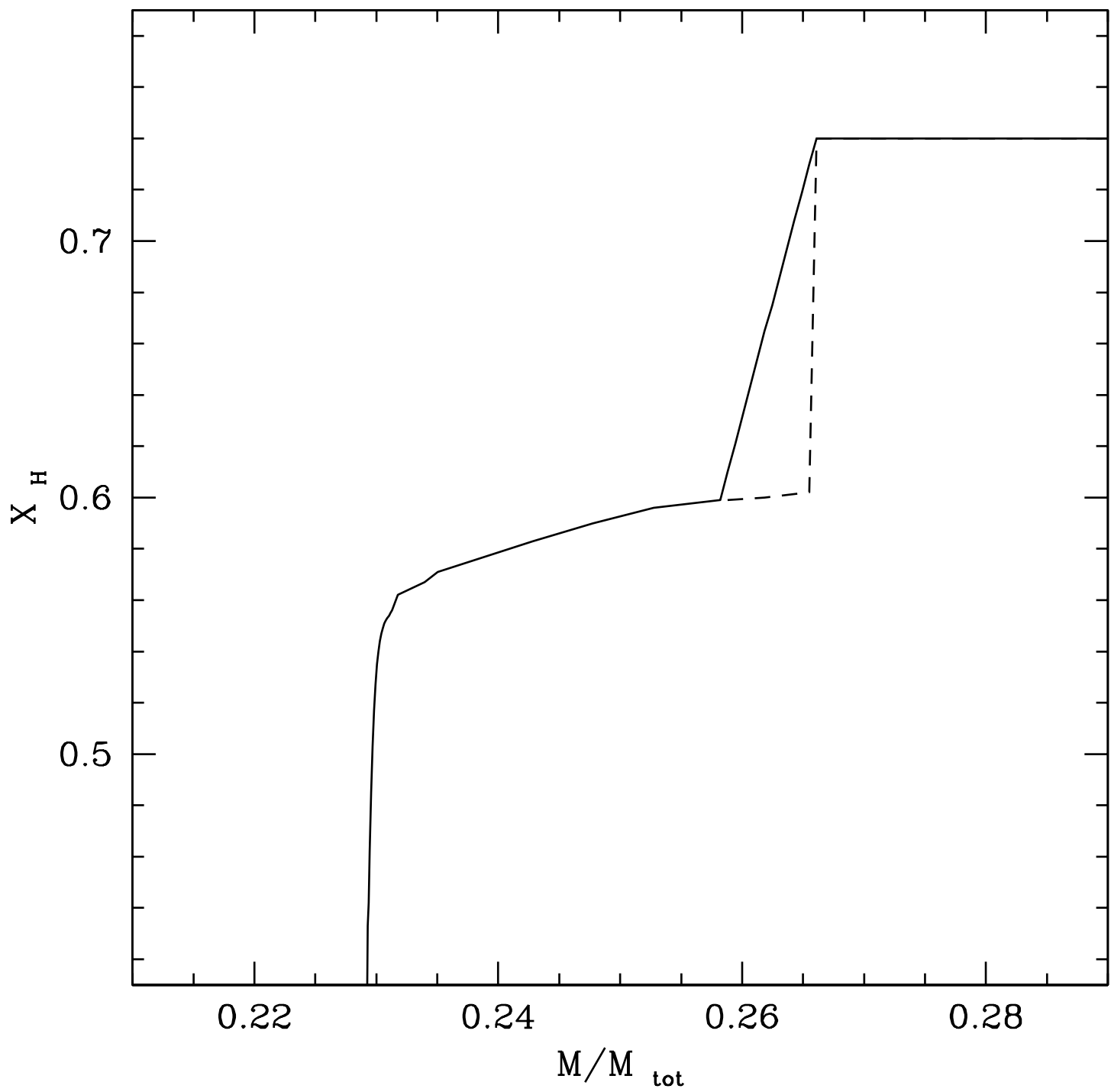
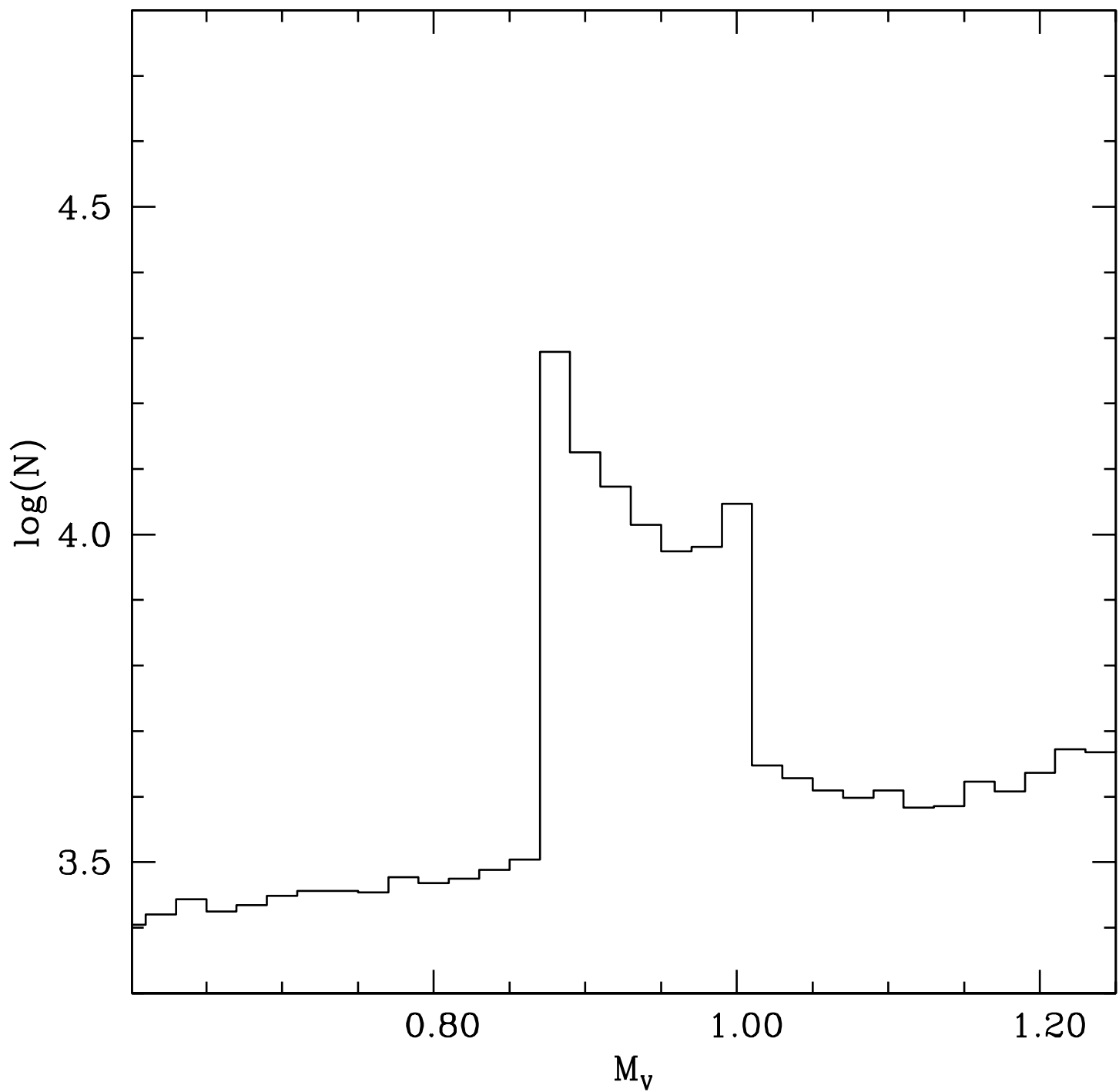
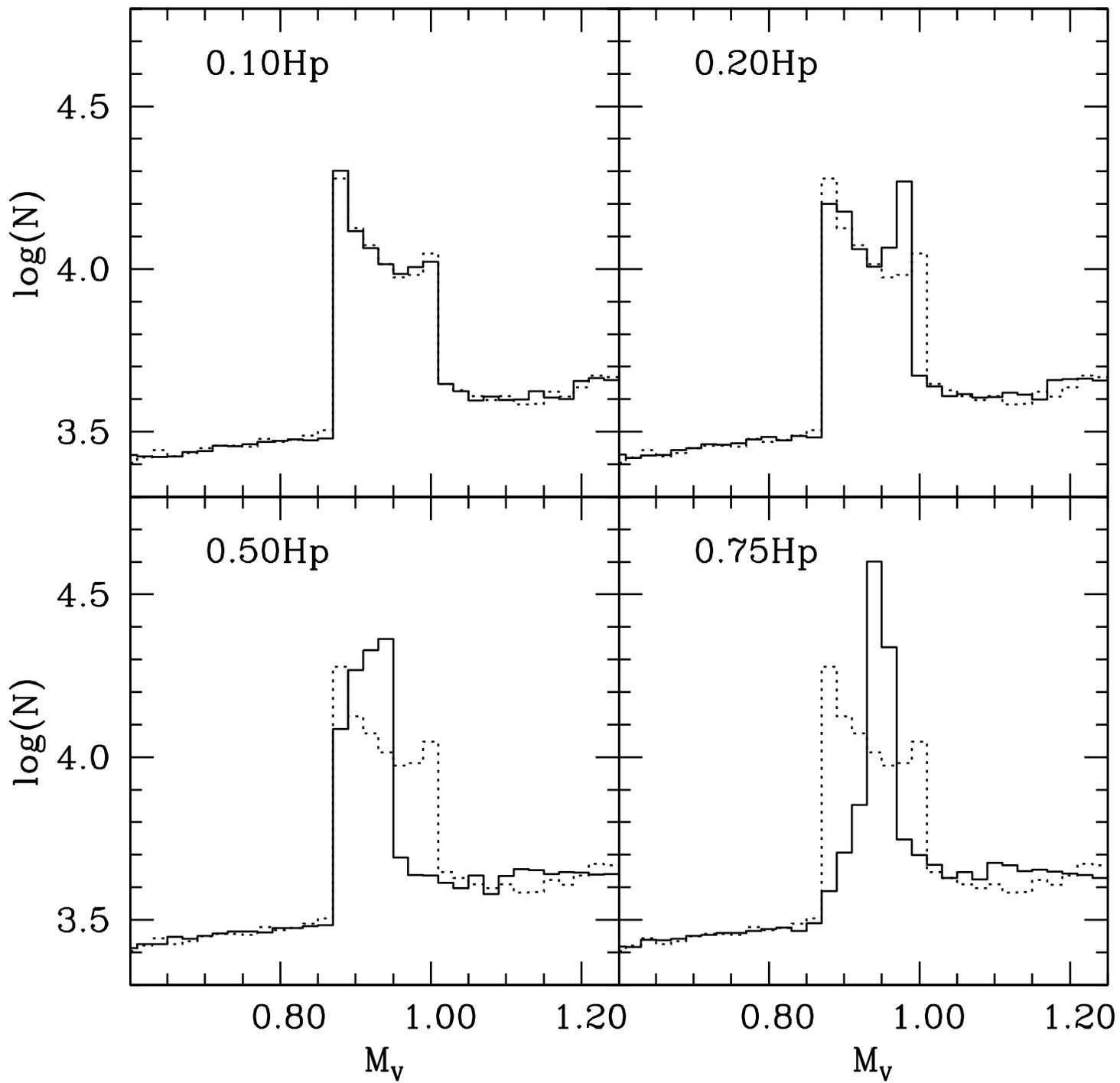
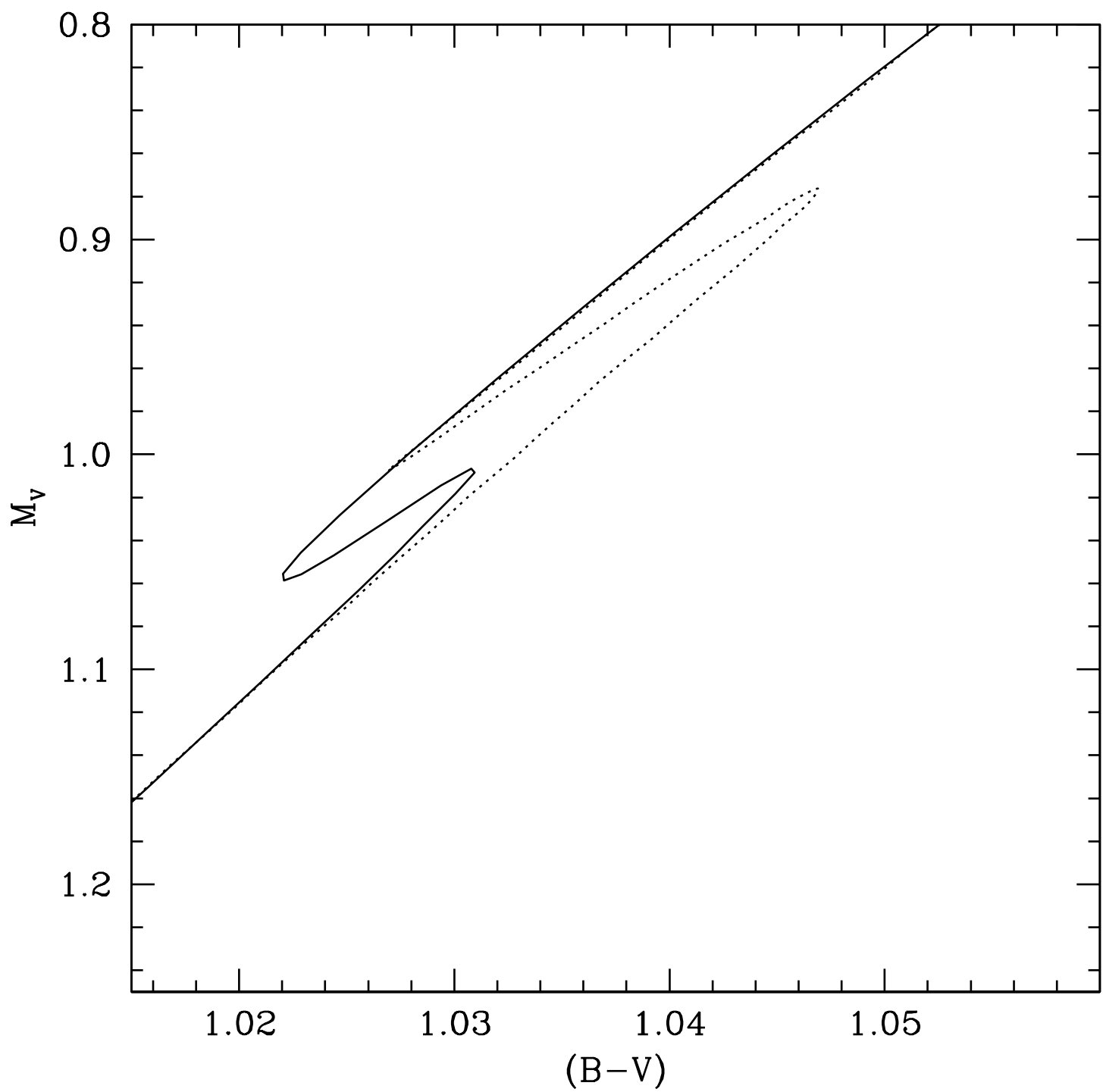
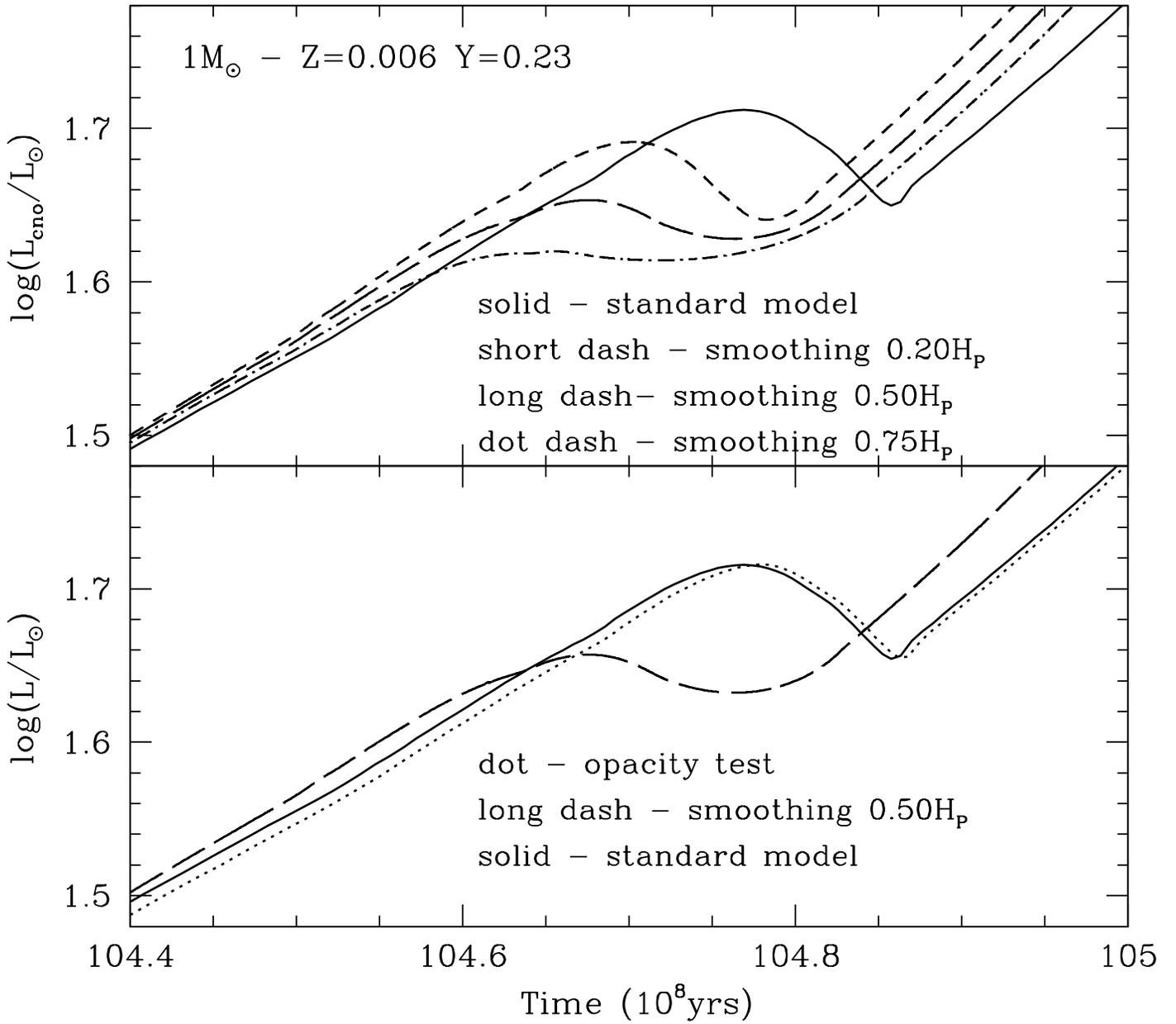
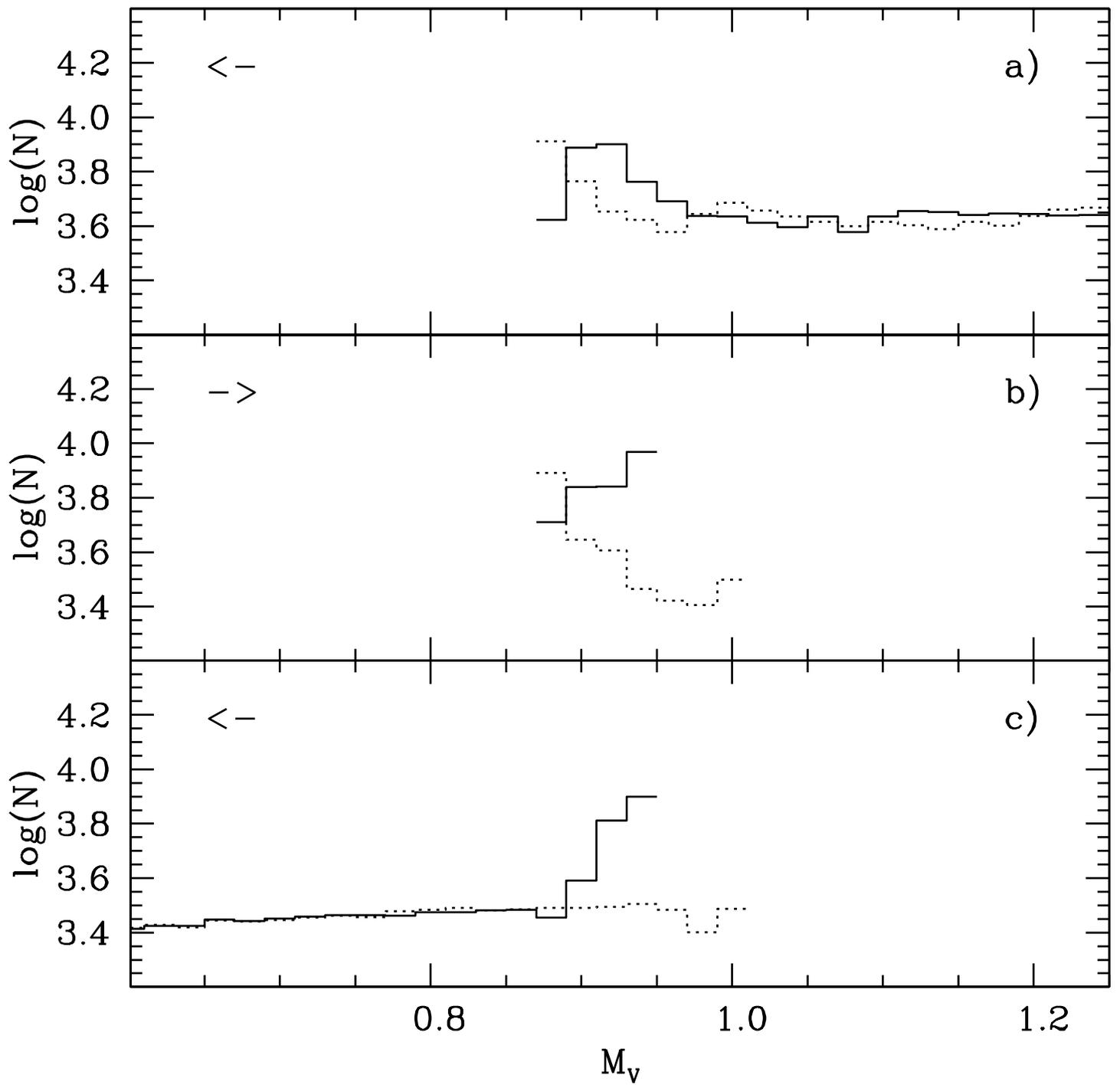
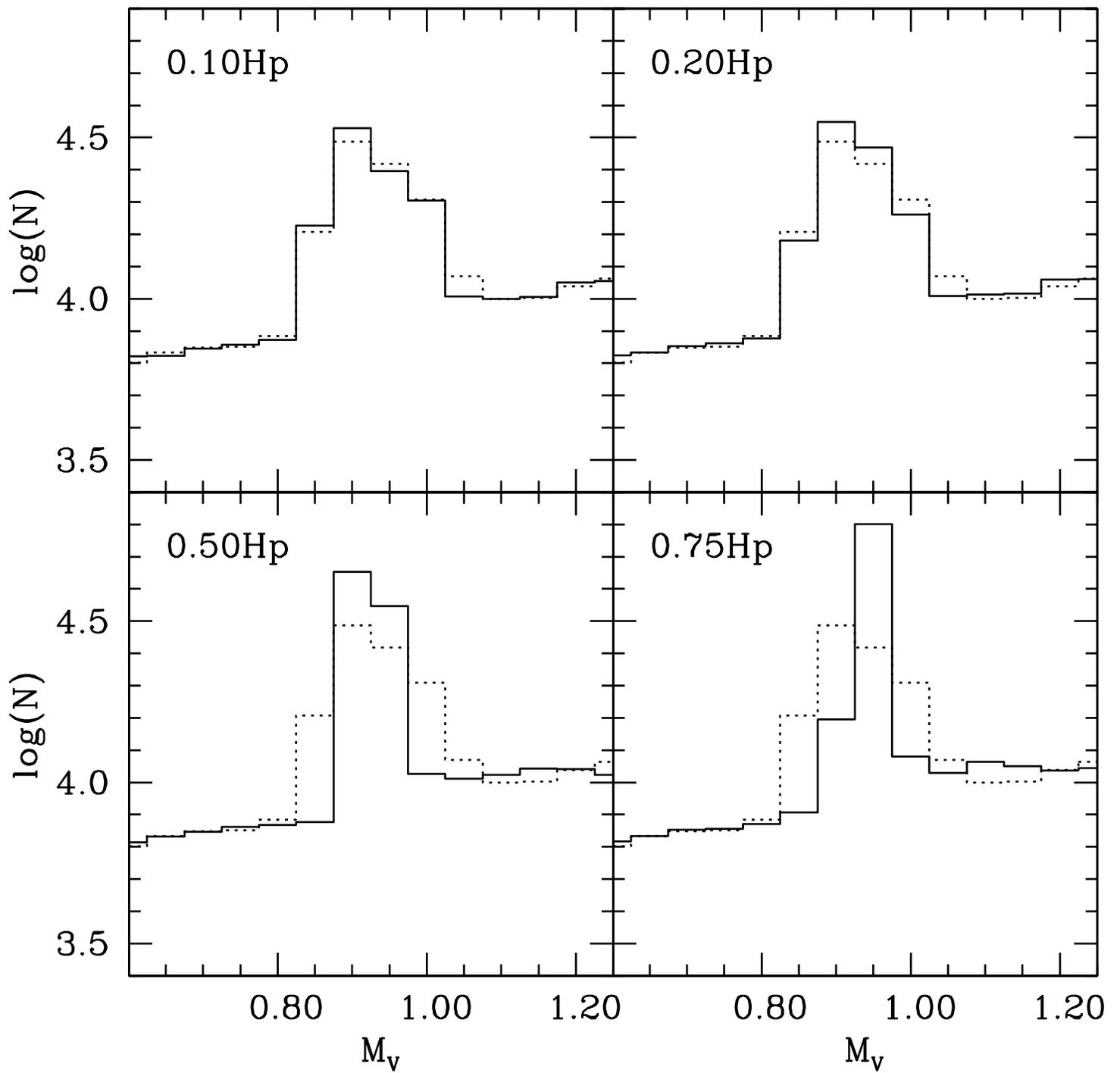
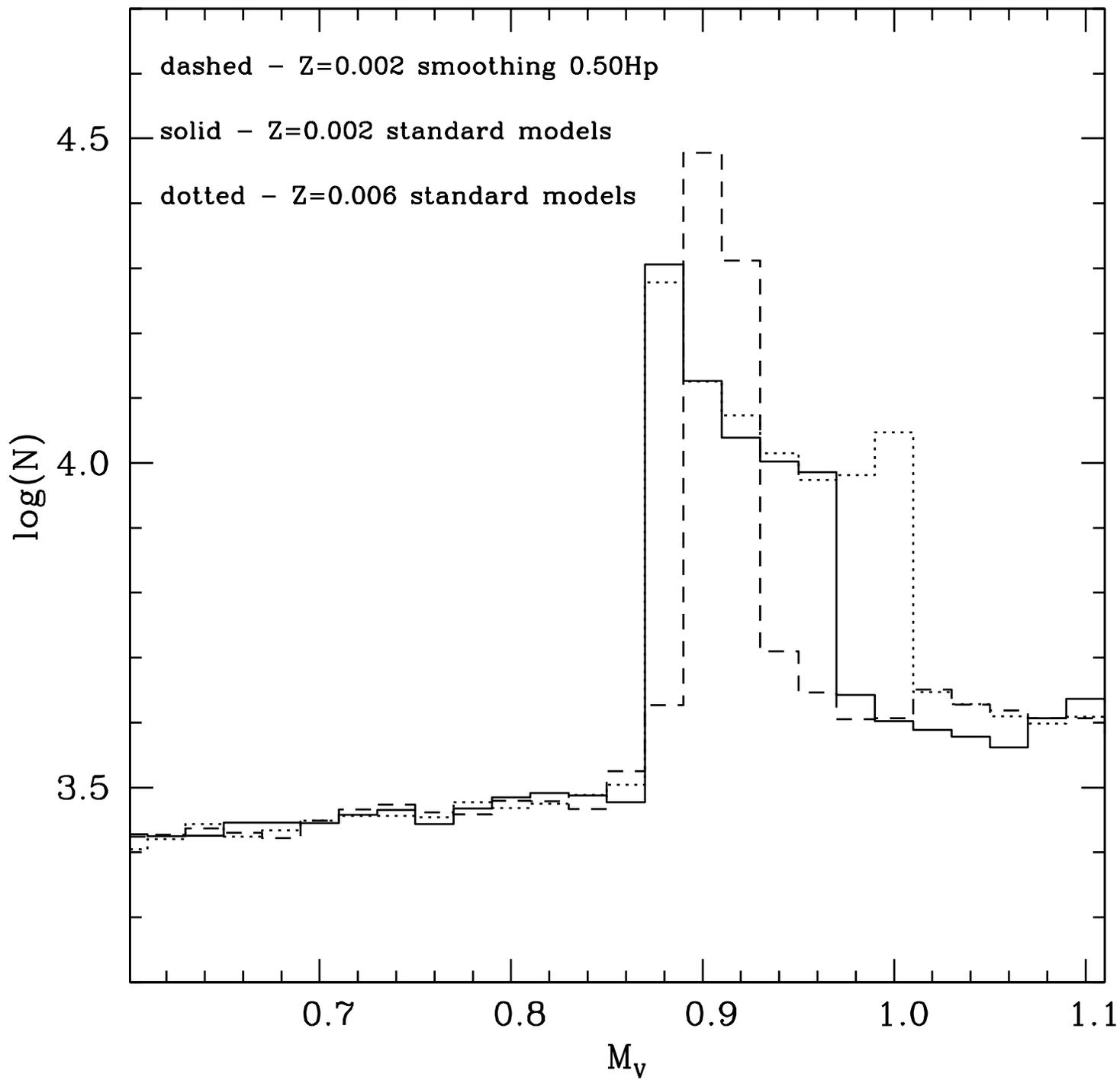
\end{document}